# PHYSICAL IDENTIFICATIONS FOR THE ALGEBRAIC QUANTITIES OF FIVE-DIMENSIONAL RELATIVITY

Paul S. Wesson[1, 2]

[1]Department of Physics and Astronomy, University of Waterloo, Waterloo, Ontario N2L 3G1, Canada

Space-Time-Matter Consortium, http://astro.uwaterloo.ca/~wesson

Addresses: Mail to Waterloo above; email: psw.papers@yahoo.ca

**Abstract**

When four-dimensional general relativity is embedded in an unconstrained manner in a fifth dimension, the physical quantities of spacetime can be interpreted as geometrical properties related to the extra dimension. It has become widespread to view the ten Einstein equations and the source terms of the energy-momentum tensor in this way. We now assign physical meanings to the other five equations involved. The scalar field acts like gravity, but concerns inertial as opposed to gravitational mass. The other four equations are conservation laws for 4D dynamics, but where the mass of a test particle is related to a local value of the cosmological 'constant'. Ways of testing these identifications are suggested.

1. Introduction

The extension of Einstein's 4D theory of general relativity to 5D as a means of unifying gravitation and electromagnetism was initially well received. But the algebra of Kaluza in 1921, as expanded to quantum effects by Klein in 1926, was restricted [1, 2], and interest waned. The theory was revived with an unrestricted algebra in the 1990s, as a means of understanding the connected problems of particle rest mass and the cosmological constant [3 – 11]. It has become a standard technique, to determine the components of the energy-momentum tensor that balance the Einstein tensor, by reducing the 5D field equations to their 4D counterparts [5]. This has enabled the discovery of many new solutions of Einstein's equations, and led to the confirmation that all 4D Friedmann-Robertson-Walker cosmologies are flat in 5D, with the implication that the big bang is due to an unfortunate choice of coordinates [10, 11]. However, such notable achievements have only directly used 10 of the 15 field equations of 5D relativity. Indeed, the main obstacle to a more universal acceptance of 5D relativity is the lack of physical interpretation of these remaining 5 equations. The theory – viewed either as an aid to general relativity or as a wider account in its own right – is physically incomplete. Despite the obvious motivation to examine these 5 field equations, giving them acceptable physical interpretations is difficult. This because we are trying to give general meanings to the rich, unrestrained algebra of 5D from the meagre physical data available on the hypersurface of spacetime in 4D. It is a bit like trying to infer the real world in ordinary 3D space from the image on a 2D photograph. In our endeavor, the main tool is covariance. The field equations of extended relativity are fully covariant in 5D, so we assume that whatever behaviour is expressed on a given 4D hypersurface is in some sense typical. In what follows, physical interpretations will be proposed for the vector and scalar components of the 5D field equations. Since our aim is to deal with *all* of the equations, some part of what follows may be familiar to certain readers. The new material joins smoothly to the old, producing a complete theory which agrees with extant observations and can be further tested as outlined at the end.

The notation is standard. Upper-case Latin (English) letters run 0, 123, 4 for time, ordinary space and the extra dimension. Lower-case Greek letters run 0, 123. The only departure from some other work is that the extra coordinate is labelled $x^4 = l$, to avoid confusion with the Euclidean coordinate and the implication that it is measured with respect to some singular hypersurface. (The latter may be inserted if desired as in membrane theory, or left out as in space-time-matter theory, reviews of which versions of 5D relativity are available in refs. 12 and 13. These two versions are mathematically equivalent, as shown and discussed in refs. 14-17.) The speed of light, the gravitational constant and the quantum of action $(c, G, h, \text{or } \hbar \equiv h/2\pi)$ will usually be set to unity, except where required for physical reasons.

2. The 5D field Equations

The field equations in 5D are commonly taken in terms of the 5D Ricci tensor to be

$$R_{AB} = 0 \quad (A, B = 0, 123, 4) \quad . \tag{1}$$

These *appear* to be relations for a vacuum, analogous to the ones $R_{\alpha\beta} = 0 \, (\alpha, \beta = 0, 123)$ which are verified by tests of general relativity in the solar system. However, (1) contain much more information. They decompose naturally into a tensor set of 10 equations, a vector set of 4 equations, and a scalar relation (see below). These come from the $R_{\alpha\beta}$, $R_{4\alpha}$ and $R_{44}$ components of (1). The tensor set is actually equivalent to Einstein's equations of general relativity, $G_{\alpha\beta} = 8\pi T_{\alpha\beta}$, *with* an effective or induced energy-momentum tensor provided by the extra terms in (1), due to the extra potential and derivatives of all the potentials with respect to the extra coordinate. That is, the 4D source is due to 5D geometry. This is nowadays understood as a consequence of Campbell's theorem, which is a result on the local embeddings of Riemannian manifolds whose dimensions differs by one [18]. Before proceeding, we wish to make some comments on embeddings which throw new light on their effectiveness.

Algebraically, it is always possible to take a 5D quantity $Q(x^\alpha, x^4; g_{\alpha\beta}, g_{4\alpha}, g_{44})$ which depends on the 5D coordinates and potentials, and split it into parts $Q_4(x^\alpha, g_{\alpha\beta})$ and $Q_5(x^4, g_{4\alpha}, g_{44}; \partial g_{\alpha\beta}/\partial x^4)$ which are purely 4D in nature or else depend on the extra dimension. If the original quantity is set to zero by field equations, like (1), then perforce we obtain a relation between the quantity $Q_4$ defined in purely 4D terms and the quantity $Q_5$ defined in 5D terms which have to do with the embedding. The former expression is *intrinsic* while the latter is *extrinsic*. For example, the curvature of the Earth can be determined either by drawing triangles (whose summed angles $\theta$, $\phi$ are not $\pi$) in the surface, *or* by boring downwards (along the radius *r*) towards the centre. Likewise, in 5D relativity the quantities determined entirely by measurements made in 4D spacetime are intrinsic, while those determined by going off the hypersurface (along the orthogonal direction $x^4$) are extrinsic. It is important to realize that while the intrinsic and extrinsic measures of a quantity may have different functional forms, they are not contradictory, but instead provide complementary measures of the same quantity. We will meet several such quantities below. While the intrinsic forms of quantities are defined in standard texts on general relativity, the corresponding extrinsic forms are new, and have to be given appropriate physical interpretations, at least on the hypersurface we call spacetime.

Returning to the field equations (1), their tensor set is a mixture of intrinsic $\left(G_{\alpha\beta}\right)$ and extrinsic $\left(T_{\alpha\beta}\right)$ components. Their physical interpretation is relatively straightforward, because the form of $T_{\alpha\beta}$ is known, particularly in the case of a perfect fluid. This has $T_{\alpha\beta} = \left[(\rho+p)u_\alpha u_\beta - pg_{\alpha\beta}\right]$, where $\rho$ and $p$ are the density and pressure, $g_{\alpha\beta}$ are the 4D potentials, and $u^\alpha \equiv dx^\alpha/ds$ are the 4-velocities defined in terms of the 4D proper time *s*. When the 4D Einstein equations are embedded in the 5D equations (1), the main gain consists of a geometrical definition for the matter parameters $\rho$ and $p$, and a deeper understanding of these. By contrast, the vector and scalar components of (1) are completely extrinsic in nature. Their algebraic form may be defined once a metric has been assigned, but their physical interpretation is open.

The field equations (1) can be written out once a form for the 5D line element is chosen. Different workers use different metrics, depending on the physical application [12, 13]. At this stage, we wish to be as general as possible, and so revert to a broad form [5]. This removes electromagnetic effects by using 4 of the 5 available degrees of coordinate freedom to set the Maxwell potentials $(g_{4\alpha})$ to zero. However, the remaining degree of coordinate freedom is held in abeyance, to bring out the effects of the scalar field $g_{44} \equiv \varepsilon \Phi^2$, where $\Phi = \Phi(x^\alpha, l)$ and $\varepsilon = \pm 1$ allows for both a spacelike and timelike extra dimension. (The extra dimension need not have the physical nature of a time even though it might be algebraically timelike, so there is no problem with closed historical paths.) The 5D line element then takes the form

$$dS^2 = g_{\alpha\beta}(x^\gamma, \ell) dx^\alpha dx^\beta + \varepsilon \Phi^2(x^\gamma, \ell) dl^2 \quad . \tag{2}$$

This includes the 4D line element $ds^2 \equiv g_{\alpha\beta} dx^\alpha dx^\beta$, where we will generally use the 4D proper time $s$ as parameter in order to make contact with extant knowledge.

With metric (2), the field equations (1) can be conveniently grouped into sets of 10 (tensor), 4 (vector) and 1 (scalar). Thus:

$$G_{\alpha\beta} = 8\pi T_{\alpha\beta}$$

$$8\pi T_{\alpha\beta} \equiv \frac{\Phi_{,\alpha;\beta}}{\Phi} - \frac{\varepsilon}{2\Phi^2} \left\{ \frac{\Phi_{,4} g_{\alpha\beta,4}}{\Phi} - g_{\alpha\beta,44} + g^{\lambda\mu} g_{\alpha\lambda,4} g_{\beta\mu,4} \right.$$

$$\left. - \frac{g^{\mu\nu} g_{\mu\nu,4} g_{\alpha\beta,4}}{2} + \frac{g_{\alpha\beta}}{4} \left[ g^{\mu\nu}{}_{,4} g_{\mu\nu,4} + \left( g^{\mu\nu} g_{\mu\nu,4} \right)^2 \right] \right\} \quad . \quad (3)$$

$$P^\beta_{\alpha;\beta} = 0$$

$$P^\beta_\alpha \equiv \frac{1}{2\Phi} \left( g^{\beta\sigma} g_{\sigma\alpha,4} - \delta^\beta_\alpha g^{\mu\nu} g_{\mu\nu,4} \right) \quad . \quad (4)$$

$$\Box \Phi = -\frac{\varepsilon}{2\Phi} \left[ \frac{g^{\lambda\beta}{}_{,4} g_{\lambda\beta,4}}{2} + g^{\lambda\beta} g_{\lambda\beta,44} - \frac{\Phi_{,4} g^{\lambda\beta} g_{\lambda\beta,4}}{\Phi} \right]$$

$$\Box \Phi \equiv g^{\alpha\beta} \Phi_{,\alpha;\beta} \quad . \quad (5)$$

Here a comma denotes the partial derivative, and a semicolon denotes the standard (4D) covariant derivative. The following three sections examine equations (3), (4), (5) by turn.

3. <u>The Tensor Equations</u>

Much has already been written about the set of equations (3) above, so the following comments are in the nature of updates. Those readers conversant with the induced-matter method may like to skip to Section 4.

(a) The Einstein tensor in (3) has its intrinsic form as specified in textbooks, while the effective energy-momentum tensor is extrinsic and comes from the extra potential $\left( g_{44} \equiv \varepsilon \Phi^2 \right)$ and derivatives of the 4D potentials with respect to the extra coordinate $\left( x^4 = l \right)$. The latter may be compared to a given physical quantity, such as $T_{\alpha\beta}$ for a perfect fluid, to identify quantities like the density $\rho$ and pressure $p$ (see above). The $T_{\alpha\beta}$ of (3) can describe all known forms of matter. When the metric (2) does not depend at all on $x^4 = l$, (3) shows $\Box \Phi = 0$, $8\pi T_\alpha^\alpha \equiv 8\pi T = 0$. This is normally interpreted to mean that the density and the (3D) averaged pressure $\bar{p}$ obey $\bar{p} = \rho/3$, the equation of state of radiation or ultra-relativistic matter. However, some care is needed here. The theory is a field description of the spin-2 graviton, the spin-1 photon and the spin-0 scaleron. All have zero rest masses, unless they acquire them by symmetry breaking (see elsewhere), and so are dynamically indistinguishable. But the waves associated with these are quadrupole, transverse and scalar, respectively. Scalar waves, in particular, have no degrees of polarization, and in that respect resemble sound waves (possibly supported, as will be discussed below, by the vacuum). Therefore, the true nature of a 5D radiation-like fluid has to be identified by looking at the source.

(b) Static solitons provide an example of this. They are radiation-like clouds, with 3D spherical symmetry which admits asymptotic flatness. However, because of the greater number of degrees of freedom in the 5D case as opposed to the 4D one, Birkhoff's theorem in its usual form breaks down. The result is a *class* of solutions of the field equations (1), associated with Kramer, Sorkin, Gross-Perry and Davidson-Owen. These authors disagree about the physical nature of the source, but the algebraic properties of its associated matter are well established [11, 19]. The perspicacious work of Ponce de Leon has in particular shown that there is a symmetry in the 5D field equations which effectively allows the first and last parts of the metric to be swapped [15, 17]. The 5D line element in quasi-Schwarzschild coordinates is given by

$$dS^2 = A^a dt^2 - A^{-(a+b)} dr^2 - A^{1-a-b} r^2 d\Omega^2 \pm A^b dl^2 \quad . \tag{6}$$

Here $d\Omega^2 \equiv (d\theta^2 + \sin^2\theta d\phi^2)$ and $A(r) \equiv 1 - 2M/r$, where we are assuming a mass-like source at the centre of the 3-geometry. The physically-dimensionless constants $a,b$ are related by the consistency relation $(a^2 + ab + b^2) = 1$. These constants $a,b$ mean that in general (6) describes a field which while fixed by the single parameter $M$ is in fact *bi-valent* in nature. To see this, let us define for use here and elsewhere the local, weak-field limit in 5D by

$$M/r \ll 1, \quad l/L \ll 1, \quad u^{123} \ll u^0 \simeq 1 \quad, \tag{7}$$

where $L$ is a length typical of the geometry and here just $M$. Then by (6), the standard procedure applied to the first and last metric coefficients identifies $aM$ as the gravitational mass of the source and $bM$ as the scalar mass. (There is an opinion in 5D relativity that the coordinate $x^4 = l$ associated with the scalar field $g_{44} = \varepsilon \Phi^2$ is a measure of the inertial rest mass of a test particle, so in some works the scalar mass of the source as just identified is also called the inertial mass of the source.) Since $a \neq b$ in general, the metric (6) is bivalent in the sense that it has gravitational and scalar contributions to the energy. Taking into account all of the metric coefficients and using a Hamiltonian approach, the total energy of a soliton is actually $(a + b/2)M$ [20]. And while attention based on 4D experience tends to focus on the first term in (6), in fact (3) shows that *all* of its associated matter outside of the central source comes from the *last* term in (6). Further, since there are no electromagnetic sources in the metric (6), the logical inference is that the cloud of radiation associated with a soliton is not photons but scalerons, or quanta of the scalar field.

(c) Time-dependent solitons help us understand why 5D objects of this type appear to be relatively rare in the universe. If the Sun, for example, is modelled by the static soliton (6), observations show that it must have $a \simeq 1$, $b \simeq 0$, meaning that its mass is nearly all gravitational and that its attendant cloud of scalerons is of negligible density [21, 22]. That is, the classic solar-system tests of relativity show that the Sun (at least) is closely described by the limit of (6), in which the 4D part is Schwarzschild and the 5th. dimension is flat. Most work on time-dependant solitons has been done for a case that is algebraically simple, with coordinates that make the 3D space $(d\sigma^2)$ isotropic. Here the extra dimension is chosen to be spacelike, and the source is relabelled $a$ (not to be confused with the constant of the preceding paragraph), and a new constant with the physical dimensions of an inverse time is labelled $H$ in analogy with Hubble's parameter of cosmology. The line element is given by

$$dS^2 = \left(\frac{ar-1}{ar+1}\right)^{4/\sqrt{3}} dt^2 - \left(\frac{a^2r^2-1}{a^2r^2}\right)^2 \left(\frac{ar+1}{ar-1}\right)^{2/\sqrt{3}} (1+Ht)d\sigma^2 - \left(\frac{ar+1}{ar-1}\right)^{2/\sqrt{3}} (1+Ht)^{-1} dl^2 . (8)$$

For this metric, (3) shows that the cloud of radiation surrounding the source at the centre of the 3-geometry is still sharply peaked, but while retaining its profile in $r$ decreases in magnitude with $t$. For $r \to \infty$ and $t \to \infty$, (8) shows that the fifth dimension disappears, and the 4D part becomes a Friedmann-Robertson-Walker model with dynamics typical of radiation. However, in accordance with preceding comments, this radiation is predicted to consist not of photons but of scalerons. Of course, what is needed here is a detailed stability analysis of the class of solitons (6), to see if the quasi-Schwarzschild case ($a = 1$, $b = 0$) is a natural endpoint of the evolution of such objects. Similarly, an analysis is needed of the effect on the formation of galaxies and other structure in the early universe if the latter contained numerous solitons [23]. The inference right now is that if 5D solitons were / are abundant, their radiation should contribute a background field of spinless particles to the matter content of the universe.

(d) The 4D Schwarzschild solution is most appropriately embedded in the 5D pure-canonical metric, not the soliton metric considered above. We will discuss the canonical metric below. It is an $l$-factorized form for the 5D line element, which leads to considerable simplification of the field equations. Any solution can be written in the form of the general canonical metric $C_5$, but only a few (with an $l$-independent 4D subspace) can be written in the form of the special canonical metric $C_5^*$. One such is the solution of (1) given by:

$$dS^2 = \frac{\Lambda l^2}{3}\left[\left(1 - \frac{2M}{r} - \frac{\Lambda r^2}{3}\right)dt^2 - \frac{dr^2}{\left(1 - 2M/r - \Lambda r^2/3\right)} - r^2 d\Omega^2\right] - dl^2 \quad . \tag{9}$$

Here $\Lambda$ is the cosmological constant and $M$ is the usual mass. There are two reasons for the belief that this is the appropriate embedding for the Schwarzschild (-deSitter) solution. *First*, it is a corollary of Campbell's theorem that any vacuum solution of the 4D Einstein equations can be embedded in a 5D metric of form (9), where the 4D metric is factorized in terms of $x^4 = l$ but otherwise independent of it, and where the fifth dimension is flat. *Second*, a test particle in the 5D field of (9) has a motion which is indistinguishable from that of the standard 4D field, a remarkable property which extends to the inflationary solution for cosmology [24]. As regards cosmology, the standard 5D models are curved with matter given by (3) in 4D, but are *flat* in 5D, where the coordi-nate transformations to Minkowski space $M_5$ are quite complicated (see e.g. ref. 11, p. 59). By contrast, the one-body metric (9) *cannot* be transformed to $M_5$. This can be appreciated either by recalling that the Schwarzschild solution cannot be embedded in a flat space of less than 6 dimensions, or by noting that the Kretschmann scalar for (9) is $K \equiv R^{ABCD} R_{ABCD} = 432 M^2 / l^4 r^6$. The constant $\Lambda$ in (9) is the intrinsic value (see above), determined by reducing the tensor set of field equations (3) to $G_{\alpha\beta} = \Lambda g_{\alpha\beta}$. It is obviously of critical importance. Despite Einstein's skeptical opinion about this parameter, it is educational to recall that his contemporary Eddington had a different view. Eddington always used $G_{\alpha\beta} = \Lambda g_{\alpha\beta}$ as the field equation for cosmology, and stated in 1932 that "To drop the cosmical constant would knock the bottom out of space" [ref. 25, pp. 22, 104]. It is somewhat ironic to learn that modern data indicate that the universe consists predominately of a $\Lambda$-like fluid [26-28], which under expansion retains its characteristic equation of state. The latter is given, with physical constants restored, by

$$p_v c^2 = -\rho_v = -\Lambda c^2 / 8\pi G \quad . \tag{10}$$

The use of this relation to implicitly include $\Lambda$ as a part of the source $T_{\alpha\beta}$ for the Einstein tensor $G_{\alpha\beta}$ has become widespread. However, it should be noted that the $8\pi G$ in the denominator of (10) exactly cancels the same term in the coupling $\left( G_{\alpha\beta} \sim 8\pi G T_{\alpha\beta} \right)$. This implies that the best way to view $\Lambda$ is the one originally due to Einstein, namely that $\Lambda g_{\alpha\beta}$ is a term that formalizes a gauge freedom of the gravitational equations. This situation is analogous to the one in classical electromagnetism as described by Maxwell's equations [29]. The situation needs to be made clear, because the field equations for 5D relativity involve another second-rank, symmetric tensor, namely $P_{\alpha\beta}$ of (4). This is distinct from $G_{\alpha\beta}$ or $T_{\alpha\beta}$, and may in principle involve another gauge term, once the algebra of (4) is given a physical interpretation.

4. The Vector Equations

The 4 field equations (4) come from the $R_{4\alpha} = 0$ components of (1). As such, they have no analogs in general relativity. However, the tensor $P_{\alpha\beta}$ exists in 4D, and it is our aim to give it a physical interpretation.

This at first appears to be difficult. In the case of the tensor equations (3) we already had a likely identification of the 5D algebra in the form of $T_{\alpha\beta}$, with a gravitational coupling provided by the weak-field limit and Poisson's equation, $\nabla^2 \phi = 4\pi G\rho/c^2$ where $\phi = GM/c^2 r$ is the potential. The physical dimensions of $G\rho/c^2$ are (length)$^{-2}$, matching the second-order derivatives of $G_{\alpha\beta}$. But $P_{\alpha\beta}$ depends on *first*-order derivatives, and has physical dimensions of (length)$^{-1}$. This has led some workers to effectively add the 'square' of $P_{\alpha\beta}$ to $T_{\alpha\beta}$ to form a composite source, the problem being still viewed as gravitational in nature. This procedure is algebraically acceptable but physically dubious, because in the form (3) the field equations $P^{\beta}_{\alpha;\beta} = 0$ show that $P_{\alpha\beta}$ is separately conserved, defining in fact a set of 4-currents. These could be electromagnetic in nature, since 5D relativity in general is a unified theory of the gravitational, electromagnetic and scalar interactions. However, the metric (2) which leads to the field equations (3) lacks electromagnetic sources. Furthermore, if the electric charge of a particle is geometrized using gravitational units, the charge/mass ratio of the particle can be written $q/m = x^e/x^m = \left(G^{1/2}e/c^2\right)\left(Gm/c^2\right)^{-1} \simeq 1\times 10^{18}$ for the proton. It is dimensionless and of enormous size for common particles, making it difficult to incorporate into theory and puzzlingly large in practice. (These problems were realized by Kaluza in 1921.) We are obliged to leave aside electric charge, admitting that it could be a parameter in $P_{\alpha\beta}$ but one whose incorporation must await an analysis based on a more general metric than the neutral-matter one of (2). And anyway, electric currents with the charges removed are still matter currents, and it is to these we now turn our attention. Even with this focus, however, there is still a question about the coupling constant for (3), because while many workers believe that 5D relativity is essentially a theory of gravitation, some believe that its 5D parts pertain to quantum physics. For example, certain individuals believe that the solitons (6) are actually magnetic monopoles [30], while others believe that the theory provides a scenario for replacing the classical big bang by a quantum event, such as tunneling in a deSitter or anti-deSitter background [31]. For versions of 5D relativity where it is applied to particle physics as opposed to gravitation, it would logically be superior to geometrize the mass $m$ of a particle by its Compton wavelength $h/mc$ rather than its Schwarzschild radius $Gm/c^2$. This is a valid issue, and we will return to it later. For now, we cut through ambiguity by proposing that the field equations (3) do indeed express the conservation of mass 'currents', and that $P^{\beta}_{\alpha;\beta} = 0$ are just the equations of motion of a test particle.

If the field equations (3) are dynamical relations, a general form for their associated tensor is

$$P^{\alpha\beta} = f_1(l) u^\alpha u^\beta + f_2(l) g^{\alpha\beta} \quad . \tag{11}$$

Here $f_1$ and $f_2$ are functions of the extra coordinate $x^4 = l$, which from the 4D perspective of a given hypersurface $(l = l_0)$ are expected to act as constants, though by preceding comments must have the physical dimensions of (length)$^{-1}$. The second part of (11) is a gauge term, analogous to the $\Lambda g^{\alpha\beta}$ which appears alongside $G^{\alpha\beta}$ in the Einstein equations, but plays no role in the Bianchi identities or conservation laws $G^\beta_{\alpha;\beta} = 0$. Accordingly, we drop the second part of (11) from our active consideration. Likewise, we regard $f_1(l)$ in (11) as a 4D constant, and concentrate on the behaviour of $^4P^{\alpha\beta} = u^\alpha u^\beta$. This may appear simple. But in the 5D theory the 4-velocity $u^\alpha \equiv dx^\alpha / ds$ is in general $u^\alpha = u^\alpha(x^\gamma, l)$ and involves the fifth dimension. Then

$$^4P^{\alpha\beta}_{;\beta} = u^\alpha_{;\beta} u^\beta + u^\alpha u^\beta_{;\beta} \quad , \tag{12}$$

and we cannot assume $u^\beta_{;\beta} = 0$ as in general relativity. To investigate this, we follow a standard technique [11]. The 4-velocities for metric (2) are normalized via

$$g_{\alpha\beta}(x^\gamma, l) dx^\alpha dx^\beta = 1 \quad . \tag{13}$$

Taking $d/ds$ of this gives

$$g_{\alpha\beta,\gamma} u^\alpha u^\beta u^\gamma + \left( \frac{\partial g_{\alpha\beta}}{\partial l} u^\alpha u^\beta \right) \frac{dl}{ds} + 2 g_{\alpha\mu} \frac{du^\mu}{ds} u^\alpha = 0 \quad . \tag{14}$$

This can be rewritten, using symmetries under the exchange of α and β and introducing the short form for the 4-acceleration. The result is

$$u^{\mu}_{;\mu} = -\frac{1}{2}\left(\frac{\partial g_{\alpha\beta}}{\partial l}u^{\alpha}u^{\beta}\right)\frac{dl}{ds} \quad . \tag{15}$$

We see that the divergence of the 4-velocity is not zero as in general relativity, because the velocities depend on the frame of reference, and there is a relative velocity between the 4D and 5D frames measured by $dl/ds$. Equivalently, there is an acceleration, or force per unit mass, which acts in 4D due to motion with respect to the fifth dimension. It is parallel to the 4-velocity, and for metric (2) is given by

$$P^{\mu} = -\frac{1}{2}\left(\frac{\partial g_{\alpha\beta}}{\partial l}u^{\alpha}u^{\beta}\right)\frac{dl}{ds}u^{\mu} \quad . \tag{16}$$

This type of force has been discussed in connection with both the space-time-matter and membrane versions of 5D relativity [32, 33]. Returning to the present analysis, putting (15) into (12) and setting the result to zero as per the field equations, there comes

$$u^{\alpha}_{;\beta}u^{\beta} - \frac{u^{\alpha}}{2}\left(\frac{\partial g_{\mu\upsilon}}{\partial l}u^{\mu}u^{\upsilon}\right)\frac{dl}{ds} = 0 \quad . \tag{17}$$

The first part of this describes geodesic motion in 4D general relativity, while the second part is a 5D perturbation.

Having gotten the motion of a test particle in 4D spacetime, it is natural to ask about the motion in the fifth dimension. For this, we put the normalization condition (13) into the metric (2) and rearrange, to give

$$S = \int \left[ 1 - \left( \frac{\Phi dl}{ds} \right)^2 \right]^{1/2} ds \quad . \tag{18}$$

Since we are using 4D proper time $s$ as the dynamical parameter, we can ask what velocity in the fifth dimension makes $S$ an extremum, and by back-substitution what this says about the 5D interval. The result is

$$\frac{dl}{ds} = \pm \frac{1}{\Phi}, \quad dS^2 = 0 \quad . \tag{19}$$

This means that particles in the manifold (2) are moving in the fifth dimension at a rate determined by the scalar field; and that as a result, all events in the manifold are in 5D causal contact.

The result $dS^2 = 0$ is of course invariant under changes of coordinates. It has been discussed in the context of space-time-matter and membrane theory [34, 35]. We can now consider rewriting the metric (2) in the form (9) of the embedded Schwarzschild solution. That is, we consider a metric of the pure-canonical form $C_5^*$, where

$$dS^2 = (l/L)^2 \overline{g_{\alpha\beta}}(x^\gamma) dx^\alpha dx^\beta - dl^2 \quad . \tag{20}$$

This has associated with it a fifth 'force' (or acceleration per unit mass) given by (16). This can be evaluated, but it is critical to recall that the 4-velocities are normalized via (13) with $g_{\alpha\beta} = (l/L)^2 \overline{g_{\alpha\beta}}$, *not* with $\overline{g_{\alpha\beta}}(x^\gamma)$. Then the force (16) results in an acceleration

$$\frac{du^\mu}{ds} = -\frac{1}{l}\frac{dl}{ds} u^\mu \quad . \tag{21}$$

This can be compared to the equivalent relation obtained from the assumption that the momentum is conserved along an *s*-path via $d(mu^\mu)/ds = 0$. This gives

$$\frac{du^\mu}{ds} = -\frac{1}{m}\frac{dm}{ds}u^\mu \quad . \tag{22}$$

By looking at (21) and (22), it is obvious that the appropriate match between algebra and physics is via $x^4 = l = m$ (or $Gm/c^2$ with units restored). In other words, metrics which can be written in the pure-canonical form (20) or $C_5^*$ have a fifth coordinate which is essentially the particle rest mass. As noted elsewhere [11], this is really not surprising, since the first part of (20) reproduces the conventional action of particle physics (*mds*) if *l* = *m*. This identification is confirmed by evaluating the constant of the motion associated with the time axis of (20), which is the energy of the test particle. It is also compatible with the fact that when the 4D metric is independent of $x^4 = l$, the resulting $T_{\alpha\beta}$ by (3) describes a radiation-like fluid consisting of particles with zero rest mass. In sum, if the coordinate $x^4 = l$ is related to rest mass *m*, 5D relativity is a theory of mechanics that treats the mechanical dimensional bases of *M, L, T* on an equal footing.

The foregoing comments about physical dimensions enable us to revisit the tensor $P^{\alpha\beta}$ as specified by (11). Specifically, now that we understand $x^4 = l$ to be related to the rest mass *m* of a test particle, we can identify the coefficient $f_1(l)$ there (though we continue to ignore the gauge term and a possible electromagnetic term). The coefficient $f_1(l)$ should clearly be proportional to *m* in order to preserve momentum in the appropriate limit; but physical constants need to be included to ensure that the physical dimensions of $P^{\alpha\beta}$ are (length)$^{-1}$. In gravitational problems, the required combination gives $f_1 = \Lambda Gm/c^2$. In quantum problems, the required combination gives $f_1 = mc/\hbar$. Re-absorbing the constants allows us to write

$$P^{\alpha\beta} = mu^\alpha u^\beta \quad . \tag{23}$$

This is the physical identification for the dynamically important part of the 4-tensor $P^{\alpha\beta}$ which appears in the vector part (4) of the field equations of 5D relativity.

5. The Scalar Equation

The scalar field equation (5) comes from the $R_{44} = 0$ component of (1). Like the vector equations considered in the preceding section, it has no analog in general relativity. However, it is in principle observable in 4D spacetime, given an appropriate physical interpretation.

Above, we saw that an argument can be made for believing that *all* test particles in 5D relativity move on 5D null-paths $(dS^2 = 0)$, even though they move on 4D paths which are null or timelike $(ds^2 \geq 0)$. Then by (2),

$$|dl|^2 = \frac{g_{\alpha\beta}(x^\gamma,l)dx^\alpha dx^\beta}{\Phi^2(x^\gamma,l)} = \phi^2 ds^2 \quad , \tag{24}$$

where $\phi \equiv 1/\Phi$ is a scalar field of the Jordan/Brane/Dicke type. The line element here is effectively changed from the standard one by the scalar field; and in the older literature there was often a discussion as to whether observations should be made in the Jordan frame or the Einstein frame. Now, it is clear that this question revolves around whether observations are made only in the 4D hypersurface of spacetime, or involve measurements of $\Phi(x^\gamma, l)$ which go off this hypersurface. That is, whether quantities are intrinsic or extrinsic, in the manner defined previously.

This issue is particularly important for the cosmological 'constant' $\Lambda$. In 4D general relativity, this is included as a gauge-freedom term in Einstein's equations and is a constant (see above). But in 5D relativity, it can be a function of the extra coordinate $x^4 = l$. To see this, let us recall that in the absence of ordinary matter, the cosmological constant is related to the 4D Ricci or curvature scalar by $4\Lambda = {}^4R$. Here, ${}^4R$ is defined in textbooks on general relativity, which give its *intrinsic* value. However, in the context of 5D relativity, there is an equally valid expression in terms of the embedding, which gives its *extrinsic* value. Thus:

$$ {}^4R = \frac{\varepsilon}{4\Phi^2} \left[ g^{\mu\nu}{}_{,4} \, g_{\mu\nu,4} + \left( g^{\mu\nu} g_{\mu\nu,4} \right)^2 \right] \quad . \tag{25}$$

This expression can be used to evaluate the *effective* value of $\Lambda$ via ${}^4R = 4\Lambda$, especially if the matter content (determined by $g_{\mu\nu,4}$) is exotic or otherwise ambiguous. A clear example is provided by the pure-canonical metric $C_5^*$ discussed previously. This embeds the Schwarzschild solution (9), and in general has the form

$$ dS^2 = (l/L)^2 \, \overline{g_{\alpha\beta}}(x^\gamma) dx^\alpha dx^\beta \pm dl^2 \quad . \tag{26}$$

Here $L$ is a constant, related to the *intrinsic* value of $\Lambda$ by $|\Lambda| = 3/L^2$. The *extrinsic* value is given by (25) as $|\Lambda| = 3/l^2$. For a spacelike extra dimension $\Lambda > 0$, while for a timelike extra dimension $\Lambda < 0$. The base-spaces are free of ordinary matter and are ac-cordingly deSitter or anti-deSitter. Algebraically, the relation between the intrinsic and extrinsic forms of a quantity is almost trivial, involving just the quadratic factor in (26): $Q_{\text{intrinsic}} = (l/L)^2 Q_{\text{extrinsic}}$. Physically, the difference is not trivial. Depending on the manner in which it is measured, it appears that data relate either to the intrinsic value $3/L^2$ (of small magnitude as determined by the dynamics of galaxies) or $3/l^2$ (of large magnitude as determined by the vacuum fields of particles). This subject is controversial, notably as regards the particle-physics side, where the scalar Higgs field is believed to be responsible for fixing masses [36-39]. Models for the Higgs field $\phi$ involve the scaling relation $\Lambda \sim \phi^2$ [38], which while arrived at by independent means is the same dependency as implied by the classical relation (25) where $\Lambda \sim {}^4R \sim 1/\Phi^2 \sim \phi^2$. The quantum and classical approaches both involve an important dependence on the scalar field.

The basic relation (5) for the scalar field has the form of a wave equation with a source. The latter depends on $\partial g_{\alpha\beta}/\partial l$, and is zero for the solitons (6). However, those solutions represent a special, static class which is picked out on physical grounds from a much broader class of oscillatory solutions, in a manner analogous to how in general relativity the Schwarzschild solution is picked out from a broader class of vacuum solutions which includes gravitational waves. Similarly, the special-canonical metric $C_5^*$ of (20) causes the first two terms on the right-hand side of (5) to cancel, leading to the solution of that equation by $\Phi$ = constant in a somewhat trivial and special way. There exist much broader classes of solutions to (3), which need investigation; especially since to any solution of (3) with $\Box\Phi \neq 0$ may be added a solution with $\Box\Phi = 0$. Such solutions may be of interest to those workers not so much concerned with classical solutions of (3) - (5) applicable to astrophysics, as wave-like solutions applicable to particle physics.

Classically, the nature of the scalar field of 5D relativity is best displayed by the solitons, whose metric (6) is an exact solution of the field equations (1), where all of the matter as given by (3) comes from the $\Phi$-field. [The pure-canonical metric (26) has only vacuum energy as specified in 4D by (10) and no 'ordinary' matter.] It was argued in Section 3 that the solitons have a source which is bivalent, in the sense that it has gravitational and scalar contributions to the energy. The parallel with electromagnetism is obvious. The question then arises of the relative strength of these terms. It is instructive to consider an 'atom' where the scalar field as per (6) dominates the electromagnetic field. Since such an object is hypothetical, the analysis is relegated to another place [40]. It shows that the effect of the scalar field on the atomic scale is much smaller than that of electromagnetism. For example, the binding energy of the orbits of test particles around a nucleus is typically less than that of the corresponding electromagnetic (Bohr) model by a factor of about 20. This implies that elementary particles bound by only the scalar field would be disrupted by some of the photons of the present-day 3K microwave background radiation. To this argument should be added the previous comment, namely that time-dependent solitons described by (8) anneal themselves over cosmic time into a background cosmological field of scalerons. It appears, therefore, that solitons do not provide a practical test for the scalar field of 5D relativity, either on the atomic or cosmological scales.

Notwithstanding this, it is relatively clear that the scalar field is connected to the masses of particles. This can be inferred from equations (21) and (22) and the comments made thereafter, which suggest that a classical definition for the rest mass of a test particle can be made via $m \equiv \int \Phi \, dl$. This is just the analog of the proper 'distance' in the fifth dimension. The dynamical properties we have discussed above are consistent with such a definition. It suggests that a massive object has both gravitational and inertial (or scalar) mass. However, the gravitational and scalar fields do not, in general, have the same source. This can be seen by combining the trace of (3) with (5). The result is an expression which gives the source for the $\Phi$-field in terms of the source for gravity $(T)$ and other terms:

$$\frac{\Box \Phi}{\Phi} = 8\pi T + \frac{\varepsilon}{2\Phi^2} \left\{ \frac{\Phi_{,4} g^{\alpha\beta}}{\Phi} g_{\alpha\beta,4} - g^{\alpha\beta} g_{\alpha\beta,44} + \frac{\left(g^{\alpha\beta} g_{\alpha\beta,4}\right)^2}{2} \right\} \quad . \tag{27}$$

The only case in which the source terms are comparable is at a matter / vacuum interface, where the discontinuity in $g_{\alpha\beta,44}$ can dominate other terms, giving $\Box \Phi / \Phi = -\varepsilon g^{\alpha\beta} g_{\alpha\beta,44} / 2\Phi^2 = -8\pi T$. Another way of evaluating the source for the $\Phi$-field is to combine (5) with the trace of (4), which is in general $P = -3 g^{\alpha\beta} g_{\alpha\beta,4} / 2\Phi$. The result is

$$\varepsilon \Box \Phi = \frac{P_{,4}}{3} + \frac{g^{\alpha\beta}_{,4} g_{\alpha\beta,4}}{4\Phi} \quad . \tag{28}$$

This is conformable with (11) and (23) for the dynamics derived from the vector set (4) of the field equations.

The net result of the considerations of this section is that the scalar field $\Phi$ of 5D relativity is considerably weaker than electromagnetism and mimics the effects of gravitation. It is therefore difficult to formulate a prognostication which might show the existence of this field in the real world. A possible test concerns the pure-canonical metric (20), which we recall predicts that $|\Lambda| = 3/l^2$ as measured extrinsically. [The scalar field $\Phi(x^\gamma, l)$ is taken to be a constant in this relation, but is implicit in the more general definition of the rest mass $m \equiv \int \Phi dl$ as outlined above.] Using gravitational units, here $l = Gm/c^2$. The result is the simple relation

$$|\Lambda|\left(\frac{Gm}{c^2}\right)^2 = 3 \quad . \tag{29}$$

That is, the magnitude of the cosmological 'constant' (as measured by the intensities of vacuum fields) and the masses of particles (as measured gravitationally) should by their combination be equal to a pure number (the value comes basically from the dimensionality of the underlying manifold). The dependency (29), while puzzling from the classical viewpoint, is what is needed to resolve the so-called cosmological-'constant' problem, which is fundamentally a mismatch between the values of $\Lambda$ inferred from particle physics and cosmology [36-39]. Provided the relevant parameters can be measured, (29) provides a test for 5D relativity. It could conceivably be carried out by the Large Hadron Collider, though the gap between concept and practice is significant.

6. <u>Summary and Discussion</u>

The 15 field equations of 5D relativity (1) with a general metric (2) split naturally into sets of 10, 4 and 1 as given by equations (3), (4) and (5). The first set reproduces Einstein's equations $G_{\alpha\beta} = 8\pi T_{\alpha\beta}$ of 4D general relativity, with a geometrical or induced energy-momentum source. Campbell's theorem guarantees such a correspondence. The second set has no Einstein analog, but the form $P^{\beta}_{\alpha;\beta} = 0$ defines 4 conservation laws, which can be related via (11), (12) and (23) to 4D laws of motion (17), which are the usual geodesic ones modified by an extra term due to the fifth dimension. The last relation of the theory is in general a wave equation for the scalar field $\Phi$, but has only been properly studied in classical contexts which show that its large-scale behaviour is like gravity, while its small-scale behaviour is largely unknown.

Another aspect of 5D relativity which is little understood concerns electromagnetism. The algebraic formulation of the theory involves definitions for the electromagnetic potentials which include the scalar field ($A_{\mu} \equiv g_{4\mu}/|\Phi|$; see e.g. ref. 11). This can lead to spatial variations of the permittivity of free space [41], modifications to the laws of geodesic motion [42], changes over cosmological times in the fine-structure constant [43], and alterations to the scattering matrix of interacting particles [44]. Electromagnetic tests of 5D relativity would be cheap to carry out, and the subject deserves further study.

The principles on which gravitational theory is based may also need re-evaluation. General relativity is commonly taken to be based on the principle of covariance, the geodesic principle and the equivalence principle. The importance attributed to these principles differs between workers, and there is an overlap between the last two. How-ever, no serious researcher would nowadays put forward a theory that was *not* covariant, so to that degree the principle is a foregone assumption of technique. As to geodesic motion, it clearly depends on the dimensionality of the manifold. Above, it was pointed out that the assumption of a 5D null-path $(dS^2 = 0)$ is natural. It matches the null nature of the field equations $(R_{AB} = 0)$. The null-path condition, it should be noted, is perfectly compatible with more general treatments of the 5D equations of motion, using either the Lagrange approach or the extremum condition $\delta(\int dS) = 0$ [11, 41]. Algebraically, $dS^2 = 0$ provides a 'short-cut' to the relative motion between the 5D and 4D frames, while the detailed motion in the latter frame is given above by the vector components of the field equations $(P^\beta_{\alpha;\beta} = 0)$. Physically, $dS^2 = 0$ means that the shortest path between events in 5D is the one which puts all particles in causal contact, irrespective of whether they are massless or massive. Viewed either way, there is no need of an explicit principle of geodesic motion in 5D, as there is in 4D.

The principle of equivalence requires more thought, as befits its fundamental status in 4D Einstein theory. However, let us consider the following line of reasoning: *Any* metric in 5D can be written in the *l*-factorized or canonical form $C_5$ given by

$$dS^2 = (l/L)^2 \overline{g_{\alpha\beta}}(x^\gamma, l) dx^\alpha dx^\beta \pm dl^2$$

$$= (l/L)^2 ds^2 \pm dl^2 \qquad . \qquad (30)$$

When here $g_{\alpha\beta} = g_{\alpha\beta}(x^\gamma \text{ only})$, we obtain the special canonical form $C_5^*$ of (20), which among other things embeds the Schwarzschild solution (9). One could even argue that in the local limit (7), the base metric of the world in 5D terms is not $M_5$ but $C_5^*$. Leaving this aside, the condition $dS^2 = 0$ (see above) leads to two types of behaviour for $l = l(s)$, depending on the signature. In terms of an arbitrary shift in $x^4 = l$ (by $l_0$) and an amplitude constant (say $l_*$), these two modes for $C_5^*$ are given by

$$l = l_0 + l_* \exp(\pm s/L) \qquad (l \text{ spacelike})$$

$$l = l_0 + l_* \exp(\pm is/L) \qquad (l \text{ timelike}) \quad . \qquad (31)$$

For both, the constant $L$ is determined by Einstein's equations in the hypersurface of spacetime by the intrinsic value of the cosmological constant $\Lambda$. (See above and refs.45, 46; for $l_0 = 0$, $\Lambda = \pm 3/L^2$ where the upper sign is for $l$ spacelike while the lower sign is for $l$ timelike; for $l_0 \neq 0$, there appears to be a divergence in $\Lambda$ which provides a formal correspondence between space-time-matter theory and membrane theory, but this lies outside the present discussion, and we drop $l_0$ henceforth.) Now in (32), $l$ plays the role of particle mass $m$, yielding thereby the standard element of 4D action $mds$. This identification was discussed in connection with the acceleration laws (21) and (22), and other supportive results were mentioned there. (For a timelike extra dimension, the observed value of $l$ is given by the root of the product of $l$ and its complex conjugate, which is just $l_*$.) However, this identification presumes that the mass is measured in gravitational units via the Schwarzschild radius $l_g = Gm_g/c^2$. This is acceptable; except that workers in particle physics may prefer to measure the mass in atomic or inertial units via the Compton wavelength $l_i = \hbar/m_i c$. The appropriate way to handle this is by the coordinate transformation on (30), $l \to L^2/l$. It should be noted that the transformed version of (30) with $dS^2 = 0$ yields $dl/ds = \pm l/L$, which is the *same* result as for the original metric and gives (31) above. (The timelike case is obtained via $l \to il$ with $L \to iL$ and $ds$ left real; the coordinate itself carries the oscillation in this case, because the scalar potential is set to unity.) The coordinate transformation involved here is $l_g l_i = L^2$. In terms of gravitational and inertial masses, this reads

$$\frac{m_g}{m_i} = \left(\frac{c^3}{G\hbar}\right) L^2 = \left(\frac{c^3}{G\hbar}\right)\left(\frac{3}{\Lambda}\right) \quad , \tag{32}$$

using the intrinsic value of the cosmological constant defined previously. This result agrees with (29). It says that the gravitational and inertial masses of a test particle are proportional to each other, which is a statement of the equivalence principle.

5D relativity is seen to be a theory which does not need to invoke two of the three principles that form the basis of 4D general relativity. The geodesic principle is replaced by the concept of a 5D null path ($dS^2 = 0$ with $ds^2 \geq 0$). The 4D equations of motion are given by the vector set $(R_{4\alpha} = 0)$ of the 5D field equations $(R_{AB} = 0)$. The principle of equivalence is replaced by a coordinate transformation, where the extra coordinate is a measure of particle rest mass when the metric is put into the canonical form. The constants of nature provide two complementary ways of geometrizing mass, ensuring that its gravitational and inertial (scalar) measures are proportional to each other. Only the principle of covariance remains, and many workers regard this as a 'given' in any case.

The theory has certain 'loose ends' that need to be addressed, and there are extra observational tests that need to be carried out. Notably, more work is needed on the defi-nition of the rest mass of a particle when the metric is *not* canonical and the extension of the vector part of the field equations to include electric charge. Observationally, certain consequences of the electromagnetic sector of 5D relativity were mentioned earlier in this section. The relationship between particle mass and the (local) value of the cosmological constant, outlined at the end of Section 5, also needs to be tested. Perhaps the most far-reaching aspect of 5D relativity, both theoretically and practically, is that the gravitational field and the scalar field are complementary consequences of the same source, notably mass. It might be difficult to admit that, in the present era of high-precision experiments, the gravitational field has to be paralleled by a scalar field. However, the situation is analogous to the components of the electromagnetic field, insofar as the electric component commonly dominates the magnetic component. In the present case, we have seen above that the scalar field may be unstable against decay to the gravitational field, and that the scalar field mimics the gravitational field. It is hoped that the new results presented here, which are consistent with previously-known ones, will provide a means of deciding if the world has only 4 or else more dimensions.

Acknowledgements

This work was supported in part by N.S.E.R.C. It includes comments made over time by several members of the Space-Time-Matter consortium, http://astro.uwaterloo.ca/~wesson.

40. Consider a hypothetical 'atom', in which the dominant force is not electromagnetism but the scalar field described by (6) of the main text. To be concrete, we assume an atomic coupling, with potential $(\hbar/Mc)r^{-1}$ where $M$ refers to the source. If $m$ refers to an orbiting test particle, for the hydrogen atom $N \equiv M/m \simeq 1835$. The balance between the centrifugal force due to the test particle's orbital velocity $v$ and the gradient of the scalar field, gives the orbital radius as $r = \hbar c/Mv^2$. If the orbital angular momentum is quantized with value $n\hbar$ where $n$ is an integer, the radius of the $n$th. orbit is $r_n = n^2 M\hbar/m^2 c = n^2 N(\hbar/mc)$. Here, the Compton wavelength of the electron is $\hbar/mc \simeq 3\times 10^{-11}$ cm, so $r_n \simeq 3\times 10^{-11} n^2 N \simeq 6\times 10^{-8} n^2$ cm for $N \simeq 1835$ (see above). The binding energy due to the scalar field for the test particle in the $n$th orbit is given as usual by integrating the force over distance, and is approximately $E_n \simeq mc^2/n^2 N^2$. For an electron orbiting a proton, the maximum-energy orbit $(n=1)$ has $E_1 \simeq 0.1\,\text{eV}$, where $r_1 \simeq 6\times 10^{-8}$ cm. If this binding energy is expressed in terms of the wavelength of a photon with the same energy, the result is approximately $1\times 10^{-3}$ cm or 10 microns. The relative strengths of the electromagnetic and scalar interactions on the atomic scale may be evaluated by taking the inverse ratio of the two typical wavelengths concerned: 5000 Å versus $10\,\mu$, which is a factor of 20.

41. Nodvik, J.S., Phys. Rev. Lett. <u>55</u>, 2519 (1985).

42. Wesson, P.S., Ponce de Leon, J., Astron Astrophys. <u>294</u>, 1 (1995).

43. Ashenfelter, T., Mathews, G.J., Olive, K.A., Phys. Rev. Lett. <u>92</u>, 041102 (2004).

44. Liko, T., Phys. Lett. B <u>617</u>, 193 (2005).